\documentclass[preprint,aps,nofootinbib]{revtex4}
                                                                                
\newcommand{\lsim}[1]{
\setlength{\unitlength}{12pt}
\begin{picture}(1.4,1.)
\put(.7,-0.3){\makebox(0.0,1.)[t]{$<$}}
\put(.7,-0.3){\makebox(0.0,1.)[b]{$\sim$}}
\end{picture}#1}
\newcommand{\gsim}[2]{
\setlength{\unitlength}{12pt}
\begin{picture}(1.4,1.)
\put(.7,-0.3){\makebox(0.0,1.)[t]{$>$}}
\put(.7,-0.3){\makebox(0.0,1.)[b]{$\sim$}}
\end{picture}#2}
\begin{document}
                                                                                
\title{Holography and Variable Cosmological Constant}
                                                                                
\author{R. Horvat}

\affiliation{
   ``Rudjer Bo\v skovi\' c'' Institute, P.O.Box 180, 10002 Zagreb,
Croatia}

\begin{abstract}
                                                                                
An effective local quantum field theory with UV and IR cutoffs correlated in
accordance with holographic entropy bounds is capable of rendering the
cosmological constant (CC) stable against quantum corrections. By setting an
IR cutoff to length scales relevant to cosmology, one easily obtains the
currently observed $\rho_{\Lambda} \simeq 10^{-47} \;\mbox{\rm GeV}^4 $, thus
alleviating the CC problem. It is argued that scaling behavior of the CC in
these scenarios implies an interaction of the CC with matter sector 
or a time-dependent gravitational constant, to accommodate the
observational data. 
\end{abstract}
                                                                                
\newpage
                          
\maketitle

It has been pointed out \cite{1,2, 3, 4} that  gravitational 
holography might provide a
natural solution to the CC problem \cite{5}. This follows primarily from the
holographic principle \cite{6} which states that the number of independent 
degrees of
freedom residing inside the relevant region is bounded by the surface area
in Planck units, instead of by the volume of the region. The principle stems
from holographic entropy bounds \cite{7, 8, 9} whose idea 
historically emerged from
the study of black hole entropy and partially from string theory. Such
bounds establish black holes as maximally entropic objects of a given size,
and postulate that the maximum entropy inside the relevant region behaves
non-extensively, growing only as its surface area.

In conventional quantum field theories the CC is not stable against quantum
corrections as there the entropy in a region of size $L$ scales extensively,
$S \sim L^3 $. Taken as a fundamental property of the microscopic theory of
quantum gravity, one can use holography to treat the CC problem. One notes
that application of the entropy bound \cite{9} to effective field theories, 
\begin{equation}
L^3 M^3 \lsim S_{BH} \sim L^2 M_{P}^2 \;,
\end{equation}
does actually suggest that an effective field theory with an arbitrary UV
cutoff $M$ must break down in an arbitrary large volume. Here the 
size of the system $L$ acts as an IR cutoff and $M_P $ is the Planck
mass. Cohen et al. \cite{2} strengthened the bound (1) considerably by
claiming that conventional quantum field theory actually fails in a much 
smaller volume. By excluding those states of the system that already have
collapsed to a black hole, they arrived at a much stringent limit
\begin{equation}
L^3 M^4 \lsim L M_{P}^2 \;.
\end{equation}
Thus, an effective local quantum field theory can be viable as a good
approximate description of physics if and only if UV and IR cutoffs are
correlated as in (2). 

One immediate implication of Eq. (2) may be seen by calculating the
effective CC generated by vacuum fluctuations (zero point energies)

\begin{eqnarray}
\rho_{\Lambda }^{ZPE} \; \propto \; \int_{L^{-1}}^{M}k^2 dk
\sqrt{k^2 + m^2} \; &\sim& \; M^4 \; \; \; \; \; \; \; \; 
M \gsim \,\,\,   m
\nonumber \\
&\sim& m M^3 \; \; \; \; \; \; 
M \lsim \,\,\,   m \;, \end{eqnarray}
since clearly $\rho_{\Lambda }^{ZPE}$ is dominated by UV modes. In both 
cases (3) the saturated form of Eq. (2) can be rewritten as 
\begin{equation}
\rho_{\Lambda }(L) \; \simeq \; L^{-2} \; G_{N}^{-1}(L) \;, 
\end{equation}
where  dependence on the IR cutoff has been made explicit not only in
$\rho_{\Lambda }$ but also in the Newton's constant, $G_{N} \equiv
M_{P}^{-2}$.\footnote{It has been argued in \cite{10} that the quartic
divergence is actually absent in $\rho_{\Lambda }^{ZPE}$ as a 
consequence of the
relativistic invariance which requires $\rho_{\Lambda } = - p_{\Lambda }$,
where $p_{\Lambda }$ is the vacuum pressure. But even so,  this has 
no influence on
the present discussion and the form of Eq. (4).} Thus Eq. (4) promotes both
$\rho_{\Lambda }$ and $G_N $ as dynamical quantities. Specifying $L$ to be the
size of the present Hubble distance $(L = H_{0}^{-1} \simeq 10^{28} \;\mbox{\rm
cm})$ one immediately arrives at the observed value for the dark energy
density
today $\rho_{\Lambda} \simeq 10^{-47} \;\mbox{\rm GeV}^4 $, provided
$\rho_{\Lambda} \simeq \rho_{\Lambda}^{ZPE}$.

Although the estimate for the CC energy density obtained by conventional means
[Eq. (3)] and supplemented by the holographic restriction [Eq. (2)] matches its
presently observed value, it has been  pointed out recently \cite{11} that  
the dark
energy equation of state is strongly disfavored by the observational data.
Namely, assuming that ordinary energy-momentum tensor associated to
matter and radiation is conserved, one easily finds using Friedman
equation that the CC and ordinary matter scale identically, $\rho_{\Lambda}
\sim \rho_m $. Hence, dark energy scales as pressureless matter $(\omega
\equiv p/{\rho } = 0)$, while the most recent data indicate 
$ -1.38 \leq \omega \leq -0.82
$ at 95\% confidence level (see e. g. \cite{12}). A proposed remedy
\cite{13, 14} of
the problem relies on the event horizon as a new choice for the infrared
cutoff $L$. In this case the present equation of state improves to $\omega
= -0.90 $; the model is, however, unable to address the cosmic `coincidence'
problem \cite{15}. 

In the present paper, we point out that taking the ordinary
energy-momentum tensor as individually quantity 
conserved may be compatible
with the possibility that $\Lambda $ be a function of the cosmological
time [as indicated by holography through Eq. (4)] only in two special cases.
In the first case, introduction of some additional terms in the Einstein
field equations is necessary; in a tensor-scalar theory of gravity, for
example, such additional terms are functions of a new scalar field. Although
this  point of view  might be welcomed in the light of the 
quintessence proposal, one should however 
bear in mind that here we deal all the time
with variable but `true' CC [Eq. (3)] with $\omega_{\Lambda} = -1 $. In
another special case, the scaling behavior of $\rho_{\Lambda}$ may be quite
different from the law $\rho_{\Lambda} \sim L^{-2}$ [Eq. (4) with $G_N $
constant], which was the basic assumption in \cite{11, 13, 14}. 

Indeed, considering the Einstein field equation
\begin{equation}
G_{\mu \nu } + \Lambda g_{\mu \nu } = 8\pi G_N T_{\mu \nu } \;,
\end{equation}
where $G_{\mu \nu } = R_{\mu \nu } - Rg_{\mu \nu }/2 $ is the Einstein tensor
and $T_{\mu \nu }$ is the energy-momentum tensor of ordinary matter and
radiation, one sees by Bianchi identities that when the energy-momentum
tensor is conserved $(\nabla^{\mu }T_{\mu \nu }=0)$, it follows necessarily
that $\Lambda = const.$. We stress that there are actually three ways to
accommodate the running of the CC with the cosmological time, $\Lambda =
\Lambda (t)$, with the Einstein field equation. The most obvious way is to
shift the CC to the right-hand side of Eq. (5), and to interpret the total
energy-momentum tensor $\tilde{T}_{\mu \nu } \equiv T_{\mu \nu } - 
\frac{\Lambda (t)}{8 \pi G}g_{\mu
\nu }$ as a part of the matter content of the universe. By
requiring the local energy-conservation law, 
($\nabla^{\mu }\tilde{T}_{\mu \nu }=0$), we obtain
\begin{equation}
\dot{\rho }_{\Lambda } + \dot{\rho }_m + 3H\rho_m (1+ \omega ) = 0 \;.
\end{equation}
We note that the time evolution of $\rho_{\Lambda }$ and $\rho_m $ is coupled
in (6). The equation of state for ordinary matter and $\Lambda $ in (6) is
$\omega $ and -1 respectively. It is important to note that both $\rho
_{\Lambda }$ and $\rho_m $ do not evolve according to the 
$\omega $-parameter from their equations of states. The important 
implication of Eq.
(6) is that there exists an interaction  between matter and the CC which
causes a continuous transfer of energy from matter to the CC and vice
versa, depending on the sign of the interaction term. The interaction
between the two components may be defined as (for pressureless matter with
$\omega =0 $)
\begin{equation}
\begin{array}{c}
\dot{\rho }_m + 3H\rho_m = X~,
\\
\dot{\rho }_{\Lambda } = -X~,
\end{array}
\end{equation}
where the coupling term $X$ is to be determined below. Taking for
definiteness $\rho_{\Lambda } = C M_{P}^{2}H^2 $, we obtain with the aid of
the Friedman equation for the flat-space case that 
\begin{equation}
\begin{array}{c}
\rho_{\Lambda } \sim \rho_m \sim a^{-3(1-\frac{8\pi C}{3})} \;,
\\
X=8\pi CH\rho_m \;,
\end{array}
\end{equation}
thus showing a rather different result for $\rho_m $ than the standard
behavior $a^{-3}$. In addition, for curved universes the scaling $\rho_m
\sim \rho_{\Lambda }$ is absent, a welcome feature for the problem of
structure formation \cite{16}.  
            
A conventional field-theoretical model with the CC running fully in
accordance with the holographic requirement (4) has been put forward
recently \cite{17}. It was based on the observation \cite{18} 
that even a `true' CC in
such theories cannot be fixed to any definite constant (including zero)
owing to the renormalization-group (RG) running effects. The variation of the CC
arises solely from the particle field fluctuations, without introducing any
quintessence-like scalar fields. Particle contributions to the
RG  running of $\Lambda $ due to vacuum fluctuations of
massive fields have been properly derived in \cite{19}, with a somewhat
peculiar outcome that more massive fields do play a dominant role in the
running at any scale. Assuming some kind of merging of quantum field
theory with quantum gravity (or string theory) near the Planck scale, one
may explore the possibility that the heaviest degree of freedom may be
associated  to particles having  masses just below the Planck scale (or
the effective value of mass in that regime 
may be due to multiplicities of such particles).
While in the perturbative  
framework of the model the running of the Newton's constant is
negligible \cite{20}, the scaling of the CC is just of the form as above,
$\rho_{\Lambda } = CM_{P}^2 H^2 $. It was shown  
(second Ref. in \cite{17}) that for $C \sim
10^{-1} - 10^{-2}$ (safely within the holographic bound) the model is 
compatible with all observational data and can be tested in future Type Ia
supernovae experiments \cite{21}. Moreover, the `coincidence' problem  is
simply understood by noting that $(\rho_{\Lambda }^0)^{1/4} 
\sim \sqrt{M_{P}H_0} $
is given by the geometrical mean of the largest and the smallness scale in
the universe today. Hence, we see that the holographic relation [Eq. (4) with
$G_N $ constant] is consistent with current
cosmological observations and does not suffer from the `coincidence'
problem. 

Let us also mention that from other considerations in line 
with the holographic conjecture, the same
law for $\rho_{\Lambda }$ has been recently reached in \cite{22}. Also,
there is a recent paper \cite{23} reaching similar conclusions from general
arguments in Quantum Field Theory. 

As already stated, if one ignores the presence of additional light scalars
(which do not influence the present discussion  anyway), the
variable CC can be achieved also by promoting the Newton's constant to a
time-dependent quantity. In this case the variation of the CC can be
maintained even if the energy-momentum tensor for ordinary matter stays
conserved. In this particular case, $\nabla^{\mu } (G(t) \tilde{T}_{\mu \nu
}) = 0 $ implies the equation of continuity to be  
\begin{equation}
\dot{G}(\rho_{\Lambda } + \rho_m ) + G\dot{\rho }_{\Lambda } = 0 \;.
\end{equation}
Hence the scaling of ${\rho }_{\Lambda }$ in (9) is coupled with the scaling
of $G$.\footnote{The more general case, of course, would have both $X \neq
0$ and $\dot{G} \neq 0$, but it is not {\em a priori } clear whether such a
model can be made compatible with the holographic reletion (4) as well as
the observational data.}

A complementary approach to that of the model \cite{17} for the RG evolution
of the CC [also obeying (9)] has been put forward in \cite{24}. Now, the RG
running is due to non-perturbative quantum gravity effects and a hypothesis
of the existence of an IR attractive RG fixed point. In contrast to the
model \cite{17}, a prominent scaling behavior of the gravitational constant
was found in \cite{24}. The behavior for the spatially flat universe was found
to be $\rho_{\Lambda } \sim H^4 $, $G_N \sim H^{-2}$. Again, the model
might explain the data from recent cosmological observations without introducing
a quintessence-like scalar field. Moreover, the model does predict that 
near the
fixed point $\rho_m = \rho_{\Lambda }$, which is quite close to the values
favored by recent observations. It is therefore up to the fixed point
structure
to provide for the mysterious approximate equality of $\rho_m $ and
$\rho_{\Lambda }$ at present (the `coincidence' problem). Hence, we  see that
our `generalized' holographic relation (4), where now both
$\rho_{\Lambda }$ and $G_N$ are varying, can be also made consistent with the 
present
cosmological data and may alleviate the cosmic `coincidence' problem.

To summarize, we have shown that the  holographic ideas discussed in the
present paper yield the behavior of the CC which is consistent with current
observations. This is true even for a `true' CC with
the equation of state being precisely -1 and with the Hubble distance as a
most natural choice for the IR cutoff.  
We have noted that the presence of
quintessence-like scalar fields is redundant in the present approach and not
required for the consistency with observational data. Our conclusion is that
the scaling of the CC stemming from holography unavoidably implies
either a 
nonvanishing coupling of the CC with dark matter or a time-dependent
gravitational constant. \newline

{\bf Acknowledgments. } The author acknowledges the support of the Croatian
Ministry of Science and
Technology under the contract No. 0098011.

\end{document}